%% file: main.tex
\PassOptionsToPackage{hyphens}{url}
\PassOptionsToPackage{svgnames,dvipsnames,x11names}{xcolor}
\documentclass[11pt]{article}

\usepackage[letterpaper,margin=1in]{geometry}
\usepackage[T1]{fontenc}
\usepackage[utf8]{inputenc}
\usepackage{microtype}

\usepackage{amsmath,amssymb,amsthm,bm}
\usepackage{cite}
\usepackage{xcolor}
\usepackage{enumitem}
\usepackage{graphicx}
\graphicspath{{figures/}{latex/figures/}}
\usepackage{booktabs}
\usepackage{multirow}
\usepackage{tabularx}
\usepackage{placeins}
\usepackage{float}
\usepackage{caption}
\usepackage{subcaption}
\usepackage{titlesec}
\usepackage{fancyhdr}
\usepackage[most]{tcolorbox}
\usepackage{hyperref}
\hypersetup{
    colorlinks=true,
    linkcolor=blue!50!black,
    citecolor=blue!50!black,
    urlcolor=blue!50!black,
    pdftitle={AtomicCommitBench: Can Coding Agents Reconstruct Commit Histories from Squashed Patches?},
    pdfauthor={Zhihao Lin, Mingyi Zhou, Li Li},
}

\titleformat{\section}{\Large\bfseries}{\thesection}{0.7em}{}
\titleformat{\subsection}{\large\bfseries}{\thesubsection}{0.7em}{}
\titleformat{\subsubsection}{\normalsize\bfseries}{\thesubsubsection}{0.7em}{}
\titlespacing*{\section}{0pt}{2.1ex plus 0.4ex minus 0.2ex}{1.0ex}
\titlespacing*{\subsection}{0pt}{1.6ex plus 0.3ex minus 0.2ex}{0.7ex}
\titlespacing*{\subsubsection}{0pt}{1.2ex plus 0.2ex minus 0.1ex}{0.5ex}

\newcommand{\findingbox}[2]{\begin{tcolorbox}
[boxsep=2pt, left=4pt, right=4pt, top=1pt, bottom=1pt,
 before skip=3pt, after skip=3pt,
 colback=yellow!8, colframe=brown!35, coltext=black, title={#1}]
#2
\end{tcolorbox}}

\title{AtomicCommitBench: Can Coding Agents Reconstruct Commit Histories from Squashed Patches?}
\author{}
\date{}

\makeatletter
\renewcommand{\maketitle}{%
  \begin{center}
    \vspace*{-1.0em}
    {\LARGE\bfseries \@title\par}
    \vspace{0.9em}
    {\large
    Zhihao Lin\textsuperscript{1}\quad
    Mingyi Zhou\textsuperscript{1}\quad
    Li Li\textsuperscript{1,*}\par}
    \vspace{0.65em}
    {\small
    \textsuperscript{1}Beihang University, Beijing, China\par}
    \vspace{0.35em}
    {\small
    \{mathieulin, zhoumingyi\}@buaa.edu.cn,
    lilicoding@ieee.org\par}
    \vspace{0.25em}
    {\small \textsuperscript{*}Corresponding author.\par}
  \end{center}
  \vspace{0.8em}
  \noindent\rule{\textwidth}{0.4pt}
  \vspace{0.9em}
}
\makeatother

\renewenvironment{abstract}
  {\begin{tcolorbox}[
    colback=gray!5,
    colframe=gray!35,
    boxrule=0.4pt,
    arc=2pt,
    left=10pt,
    right=10pt,
    top=8pt,
    bottom=8pt]
   \small\noindent\textbf{Abstract.}\enspace}
  {\end{tcolorbox}\vspace{0.6em}}

\begin{document}

\maketitle

\begin{abstract}
Coding agents often finish a session by returning one squashed patch that mixes feature implementation, bug fixes, refactorings, tests, and configuration edits. However, implementing the requested functionality is not enough: when unrelated edits are collapsed into one patch, the history no longer records the intentions, dependencies, and review units that support later maintenance. Human maintainers and coding agents that rely on repository history must infer this missing structure from the final diff alone. Since coding agents already reason over code and diffs during development, a natural question is whether they can also maintain history retrospectively. We study \emph{retrospective commit-history reconstruction}: given a completed squashed change, an agent groups its hunks into commits and materializes them as a replayable sequence. We formalize the task as hunk-to-commit partitioning with a replay requirement and build AtomicCommitBench from 800 real consecutive-commit episodes across 10 Python projects. Because developers can decompose the same change in more than one reasonable way, we score outputs with PPAR for replay validity, ARI for reference-based grouping, and TCR for failure containment on scoreable modified-test episodes. Natural retrospective reconstruction exposes difficulty that replay checks and synthetic tangling miss. All non-random methods nearly saturate replay validity (PPAR $\geq 0.988$), yet grouping quality ranges from 0.03 to 0.46 ARI. Matched synthetic composites are much easier for an untangling baseline than real same-author squashed diffs (+0.333 ARI). In this model-agent snapshot, the GPT-5.4 setup driven by Codex CLI (0.46) and the GLM-5 setup driven by Claude Code (0.43) form a higher ARI band than the MiniMax (0.31) and Kimi (0.29) setups. On the 151 episodes scoreable by the partial modified-test probe, TCR gives the same setup ordering without using developer labels. These signals indicate that current setups can produce replayable, structured draft histories with meaningful alignment to human-maintained organization, with the strongest setups recovering structure beyond file-local grouping. Qualitative trace diagnosis highlights same-file lumping and support-hunk drift as recurring error patterns. Dependency-Aware Commit Evidence (DACE) improves the two lower-scoring setups (+0.05 to +0.08 ARI), consistent with dependency cues and hunk roles helping when agents over-rely on locality. AtomicCommitBench lets agentic-coding evaluation measure the history an agent leaves behind alongside the final code.
\end{abstract}

\input{sections/introduction}

\input{sections/formalization}

\input{sections/dataset}

\input{sections/evaluation}

\input{sections/related}

\input{sections/discussion}

\input{sections/threats}

\input{sections/conclusion}

\FloatBarrier
\bibliographystyle{IEEEtran}
\bibliography{references}

\end{document}

%% file: sections/introduction.tex
\section{Introduction}
\label{sec:intro}

Coding agents are moving software work from edit-by-edit programming toward session-level change production. In one session, an agent may explore candidate fixes, touch many files, add tests, refactor nearby code, and update configuration before returning a final patch. Human-in-the-loop use can make the scope even less linear: developers steer agents with follow-up requests, parallel feature requests, or partial corrections instead of waiting for one cleanly scoped change. The resulting patch may implement the requested functionality, yet still leave a history that is hard to inspect or reuse.
Current coding-agent benchmarks~\cite{jimenez2024swebench,chowdhury2024swebenchverified,zan2025multiswebench,zhang2025swebenchlive} evaluate final-patch correctness and leave history organization unmeasured.

After a coding session, teams rely not only on the final code but also on the history that explains it. A monolithic commit and a clean sequence of atomic commits may contain the same diff, yet they behave differently under review, \texttt{git bisect}, selective revert, and history-based retrieval. A bug fix tangled with a refactor costs more reviewer attention. A version bump bundled with a behavior change is harder to revert safely. The retrieval case is concrete as well. A single feature rarely consists of one edit in one place: it may require a new subclass, a config registration, a type stub, and a test. Distributed changes like this are easy to ship incomplete. The gap often surfaces later, after a user hits it and a follow-up commit supplies the missing piece. Clean, self-contained commits let the next developer or agent recover the complete pattern. If the fix sits inside an unrelated commit, the recoverable example may be the original incomplete change, so the same bug can reappear.

Prior software engineering results already tie commit structure to later work. Decomposed changesets can reduce wrongly reported issues and context-seeking during review~\cite{dibiase2018change}; tangled changes can inflate defect prediction error by 5 to 200\%~\cite{herzig2016impact}. Recent agent pipelines also read commit history: Code Researcher~\cite{singh2025coderesearcher} searches repositories using commit history, HAFixAgent~\cite{shi2025hafixagent} conditions repair on commit-level context, and Lore~\cite{stetsenko2026lore} treats commit messages as a structured protocol. Histories produced by today's agents can become inputs for later agents. We therefore treat commit organization as a measurable producer-side property of an agent's output.

Good history organization is more than partitioning a diff by file or label. Real development already interleaves change intents: behavior edits may sit beside same-file refactoring, tests, release metadata, or configuration updates. Some change intents cross files, while others split hunks within one file. In our 800-episode sample from GitHub projects, an episode contains a median of 12 hunks across 6 files, with a mean of 51 hunks across 9.8 files. In 35.4\% of episodes, at least one file contains hunks from different observed commits; the Hard tier reaches 55.0\%. Meanwhile, 59.5\% of observed commits span at least two files. A method has to separate nearby hunks when they serve different change intents and keep distributed hunks together when they implement one logical change.

Therefore, we study the problem as \emph{retrospective commit-history reconstruction}. The input is a completed squashed change and the repository snapshot before that change. The agent organizes a known diff into commits without generating new code or solving a hidden issue. This setting isolates history organization from patch construction and matches common cleanup workflows, such as splitting a squashed branch or preparing a large pull request for review. To obtain observable references for this cleanup setting, we use human-maintained consecutive commit histories: the original commits give one defensible organization of a completed change, while the agent sees only the squashed diff and base snapshot. The central prediction is a hunk-to-commit grouping: the output assigns each hunk to one predicted commit and chooses the number of commits. The groups are then sequenced so that the result is a replayable commit history.

\begin{figure}[H]
\centering
\includegraphics[width=0.95\linewidth]{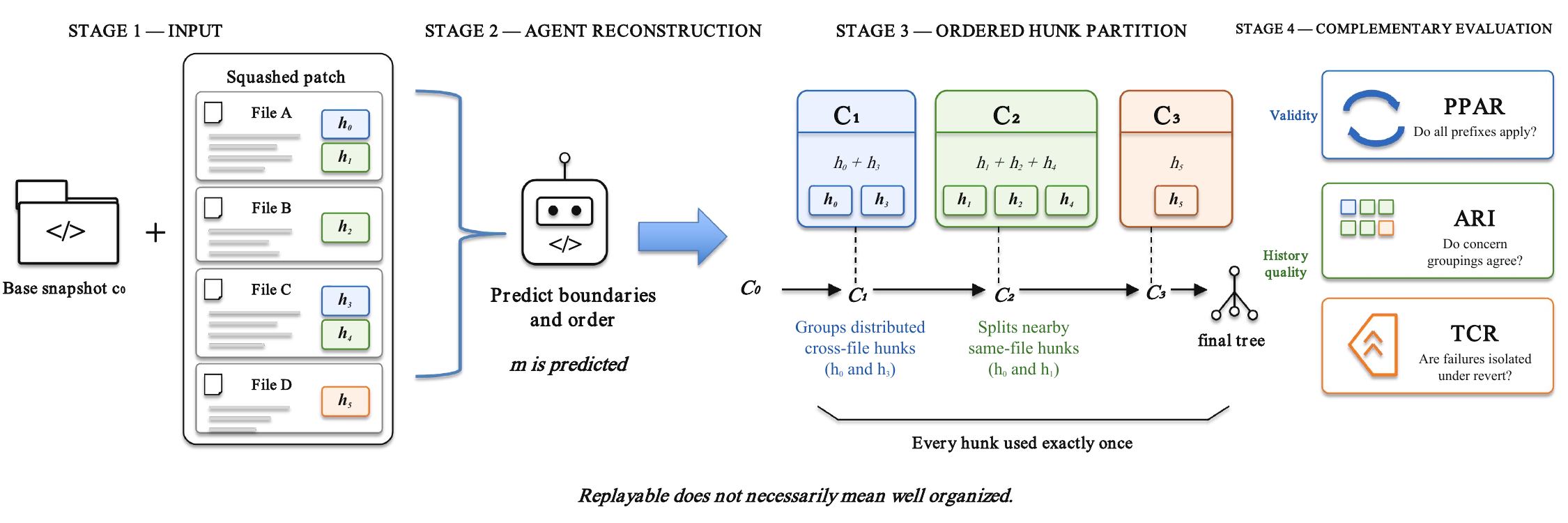}
\caption{Overview of retrospective commit-history reconstruction. Given a base repository snapshot and a squashed multi-commit diff, the task is to group hunks into commits and materialize a replayable history. Evaluation separates structural replay validity, reference-based grouping quality, and selective-revert failure containment.}
\label{fig:overview}
\end{figure}

This task has no single correct decomposition. Reasonable developers may group small cleanups differently or attach support edits to different behavior changes. We avoid a hard semantic-validity rule and ask three narrower questions: whether the predicted history replays, how its hunk grouping compares with an observed developer decomposition, and whether selective-revert failures stay contained within predicted commits. One observed developer history serves as a reference rather than the unique ground truth.

AtomicCommitBench contains 800 real multi-commit episodes from 10 Python projects. The public artifact is available at \url{https://github.com/mathieu0905/AtomicCommit}. We evaluate four model-agent setups against six diagnostic baselines. We also compare real squashed changes with matched synthetic composites to test whether the synthetic tangles used in prior work capture the same boundary structure. We organize the study around three questions:

\begin{enumerate}
    \item \textbf{RQ1: What makes natural retrospective reconstruction non-trivial?} We test whether replay validity, simple heuristics, and synthetic composites capture the same difficulty as real squashed changes.
    \item \textbf{RQ2: How do current model-agent setups perform under complementary history-quality signals?} We compare four model-agent setups using replay validity, reference-based hunk grouping, and a partial selective-revert containment probe.
    \item \textbf{RQ3: What failure modes explain the remaining gap, and when does explicit evidence help?} We qualitatively analyze same-file lumping and support-hunk drift, then test whether Dependency-Aware Commit Evidence (DACE) exposes useful dependency and hunk-role cues.
\end{enumerate}

The paper makes three contributions:
\begin{enumerate}
    \item \textbf{A formalization of retrospective commit-history reconstruction.} We turn the organization of a completed change into a hunk-level commit decomposition task, separating history quality from final-patch correctness.
    \item \textbf{AtomicCommitBench and a measurement design for non-unique histories.} The benchmark contains 800 real episodes from 10 Python projects and scores outputs with structural replay validity (PPAR), reference-based grouping quality (ARI), and selective-revert failure containment (TCR). This design uses the observed developer history as a reference without treating it as the only valid answer.
    \item \textbf{Empirical evidence about difficulty, current agents, and failure mechanisms.} Real same-session squashed changes are substantially harder than matched synthetic composites for an untangling baseline. Replay validity nearly saturates while grouping quality varies, and DACE helps mainly where agents over-rely on locality in this setup.
\end{enumerate}

%% file: sections/formalization.tex
\section{Formalization}
\label{sec:formalization}

\subsection{Retrospective Commit-History Reconstruction}
\label{sec:retrospective}

To evaluate history organization independently of code generation, we fix the final change and ask only how it should be organized. A clean history can be produced in two ways: an agent may commit incrementally while writing the code, or it may reorganize the completed change after the code is finished. We study the second setting because incremental commits can be entangled with interactive problem solving: the agent may misunderstand the developer's intent, implement a wrong direction, receive corrections, and backtrack. A messy incremental history may therefore reflect miscommunication and repair during the session rather than the agent's ability to organize the completed change. Starting from a completed squashed change removes this confound. In retrospective commit-history reconstruction, the agent receives the squashed diff and the repository snapshot immediately before it. The code change is fixed; the agent does not generate new code or solve a hidden issue. Its task is to group the diff into coherent commits that can be replayed for review, selective revert, and later reuse.
Maintainable decompositions sometimes split hunks within one file and sometimes group hunks across files. The natural unit of prediction is therefore the hunk. We formalize the task as a hunk partition with a replay requirement.

\subsection{Hunk Partition and Replay Task}
\label{sec:task}

\textbf{Definition 1 (Change Episode).} A change episode $E=(R,c_0,(c_1,\ldots,c_k))$ consists of a repository $R$, a base commit $c_0$, and $k$ consecutive commits by the same author, where each $c_i$ applies cleanly on $c_{i-1}$. The episode induces a squashed diff $D$ from $c_0$ to $c_k$, and the original commits provide one observed decomposition of that diff.

\textbf{Definition 2 (Hunk Partition with Replay).} Given the squashed diff $D$, we parse it into hunks $H=\{h_1,\ldots,h_n\}$. Each hunk is a maximal contiguous block of added and deleted lines, the granularity at which tools such as \texttt{git add -p} stage changes. At repository state $c_0$, the method partitions $H$ into a sequence $\pi=(C_1,\ldots,C_m)$ of non-empty groups; each group $C_i$ is a predicted commit. The predicted number of commits $m$ is chosen by the method and need not equal the observed number $k$. A valid output must satisfy:
(1)~\textbf{Coverage}: $\bigcup_i C_i = H$ and $C_i \cap C_j = \emptyset$ for $i \neq j$;
(2)~\textbf{Replay Validity}: applying the groups in order produces a chain of intermediate states $s_0 \to s_1 \to \cdots \to s_m$, where $s_0$ is the repository state at $c_0$ and each $s_i$ results from applying the hunks in $C_i$ to $s_{i-1}$ without conflicts.
By coverage, the final state $s_m$ equals the post-change state $c_k$, so a valid output is exactly a commit history that replays the squashed change from $c_0$ to $c_k$.

We keep semantic coherence out of the validity definition. In an ideal history, each commit should address one coherent change intent, but developers can draw those boundaries differently~\cite{herzig2016impact,herbold2022tangling}. One developer may keep a small cleanup with the bug fix that motivated it; another may separate it. We measure semantic coherence empirically instead of encoding one developer's style as a rule. Under this definition, a one-commit output can be formally valid and still be a poor decomposition.

\subsection{Relation to Commit Untangling}
\label{sec:untangling_relation}

The closest prior task is commit untangling, which groups a tangled change by change intent~\cite{shen2021smartcommit,barnett2020flexeme,wang2022utango}. That line of work casts the problem as grouping against a reference partition. Commit-history reconstruction uses the same central intuition, but the output is a commit-history artifact rather than a partition alone: after hunks are grouped, the groups must be sequenced so their prefixes replay cleanly through valid intermediate states. Reviewers inspect one commit at a time, \texttt{git bisect} searches over commit prefixes, and \texttt{git revert} removes whole commits from a history. We therefore keep grouping quality primary and use replay validity as the structural condition that turns a partition into a history.

%% file: sections/dataset.tex
\section{Methodology}
\label{sec:methodology}

\subsection{Dataset Construction}
\label{sec:dataset_construction}

We construct the evaluation set from 10 mature Python projects on GitHub. The projects cover testing (pytest), data validation (pydantic), HTTP clients (requests and httpx), formatting (black), terminal rendering (rich), web frameworks (fastapi and flask), and command-line interfaces (click and typer). We chose popular projects with mature public histories and active development. The observed histories from these mature projects provide generally high-quality human decompositions of real changes. They serve as reference decompositions for evaluating whether agents can reconstruct maintainable histories from squashed changes. Individual commits may still reflect developer style or non-atomic practice, so we do not treat any observed decomposition as uniquely correct.

We collect candidate episodes with PyDriller~\cite{spadini2018pydriller} from repository snapshots taken between 2025-12-30 and 2026-03-09. For each project, we scan the commit history for maximal same-author runs in which adjacent commits are at most 48 hours apart. We keep runs with 2 to 8 commits and discard longer runs rather than splitting them into sub-windows. We exclude runs that contain merge commits, bot-authored commits, documentation-only changes, binary files, diffs smaller than 20 lines, or diffs larger than 2{,}000 lines. A run enters the candidate pool only if its commits replay cleanly from the base commit and the replayed final state matches the squashed diff. This process yields 2{,}558 candidate episodes before sampling, with commits dated from 2007 to 2026. The distribution is uneven: pytest (921) and pydantic (610) contribute the most, while click (55) and typer (24) contribute the fewest. Each accepted episode is squashed into one unified diff with Git's default three lines of context and parsed into hunks. We assign each hunk to the commit contributing most of its lines, producing reference labels used only by the evaluator. Agents receive only the squashed diff and base repository snapshot; they do not receive commit messages, commit hashes, intermediate states, or original commit boundaries.

The same-author and time-window constraints approximate coherent development sessions without requiring issue metadata. They reduce the risk of joining unrelated work by different contributors or long-lived branches spanning multiple tasks. This filtering favors retrospective cleanup episodes, matching the setting studied in this paper.

\subsection{Difficulty Stratification}
\label{sec:difficulty}

We stratify episodes by commit count and diff size. \textbf{Easy} episodes have at most 3 commits and 200 changed lines (1{,}461 episodes). Among the remaining episodes, \textbf{Medium} episodes have at most 5 commits and 600 changed lines (415 episodes), and the \textbf{Hard} tier contains the remaining 682 episodes. We then sample 800 episodes with seed 42: 320 Easy, 320 Medium, and 160 Hard. The sample oversamples Medium cases so that the comparison includes enough non-trivial episodes.

\begin{table}[t]
\centering
\caption{Sample composition and per-episode statistics by difficulty tier. Hunks and files are reported as mean (median); percentage rows report rates.}
\label{tab:episode_stats}
\small
\begin{tabular}{@{}lrrrr@{}}
\toprule
Metric & Easy & Medium & Hard & All \\
\midrule
Episodes & 320 & 320 & 160 & 800 \\
\addlinespace[2pt]
Observed commits & 2.33 & 3.47 & 5.07 & 3.33 \\
Changed lines & 68 & 260 & 701 & 272 \\
Hunks & 19 (6) & 71 (19) & 78 (31) & 51 (12) \\
Files & 4.8 (4) & 10.3 (7) & 18.8 (13) & 9.8 (6) \\
\addlinespace[2pt]
Intra-file tangling & 22.2\% & 38.8\% & 55.0\% & 35.4\% \\
Multi-file commits & 57.0\% & 60.0\% & 63.6\% & 59.5\% \\
\bottomrule
\end{tabular}
\end{table}

Table~\ref{tab:episode_stats} reports the sampled composition and per-episode statistics by tier. Harder tiers have larger diffs, more files, and more intra-file tangling. Multi-file commits are common even in Easy episodes, so file-level grouping misses cross-file structure across the whole sample.
The difficulty tiers are used for analysis, not for scoring. Every method is evaluated on the same 800 episodes, and the headline ARI is the unweighted mean over this fixed evaluation set; re-weighting to the original pool proportions does not change the setup ranking.

\subsection{Metrics}
\label{sec:metrics}

The formal task uses hunks as the prediction unit. A hunk could in principle contain lines from more than one original commit, which would make hunk-level prediction lossy. Auditing the squashed hunks against the per-commit reference labels shows this is rare: 97.7\% draw all of their lines from a single observed commit, and the remaining 2.3\% are cases where hunk-level prediction is lossy.

Because the same change can admit multiple defensible decompositions, evaluation needs signals that do not reduce the task to exact recovery of one observed history. We combine three complementary measurements. Structural replay checks whether the predicted sequence is executable, reference-based grouping compares the hunk partition with one human-maintained history, and selective-revert failure containment tests whether predicted commits localize behavioral breakage without using developer labels.

\textbf{PPAR} (prefix-patch apply rate) measures structural replay validity. For a predicted sequence $\pi=(C_1,\ldots,C_m)$, we apply each prefix $(C_1,\ldots,C_i)$ and compute the fraction of prefixes that apply without conflicts. PPAR directly uses the predicted order. It rejects non-replayable prefixes and stays silent about which replay-valid order is more reviewable.

\textbf{ARI} measures grouping quality with Adjusted Rand Index~\cite{hubert1985comparing} between the predicted hunk partition and the observed developer partition. ARI ignores commit names, index labels, and order, so it measures whether hunks are grouped together while leaving replay validity to PPAR. Because it uses one observed decomposition as reference, we interpret it as a dataset-level relative signal rather than an absolute per-episode correctness score.

\textbf{TCR} (test-failure containment rate) is a label-free probe of whether predicted commits behave like isolated maintenance units. For episodes that modify executable Python tests, we first run the modified tests at the episode tip and keep the tests that pass. We then evaluate each predicted commit by applying only that commit's reverse patch to the tip state and rerunning the retained tests. TCR is the fraction of induced test failures that appear under exactly one predicted commit's reverse patch. Intuitively, if a logical change is split across commits or unrelated changes are lumped together, failures tend to spread across multiple reverse patches rather than localize to one commit. An episode is scoreable only when the modified tests run at the tip and at least one retained test fails under a partial reverse patch; 151 common episodes meet these conditions in the comparison between agent outputs and the observed history. Episodes outside this slice remain in the PPAR and ARI evaluation. We pair TCR with \textbf{TBR} (test breakage rate), the fraction of retained tests broken by a partial reverse patch, because containment is meaningful only when it does not come from broad breakage.

\subsection{Diagnostic Baselines and Evidence-Augmented Runs}
\label{sec:baselines}

We use six heuristic baselines to diagnose what simple organization rules can and cannot capture. \emph{NoSplit} (B0) places all hunks in a single commit and measures how much score a vacuous replayable answer earns. \emph{Random} (B1) partitions hunks randomly into two to five commits. \emph{FileSplit} (B2) creates one commit per changed file, testing a common cleanup shortcut. \emph{HunkSplit} (B3) creates one commit per hunk. \emph{DepSplit} (B4) groups hunks by syntactic dependency components. \emph{Untangle} (B5) vectorizes path and hunk tokens with TF-IDF, clusters hunks with agglomerative clustering, and orders clusters with a dependency-based topological pass. It adds lexical similarity and a lightweight ordering repair while leaving program semantics and change intent unmodeled.

For the failure-mechanism analysis in RQ3, we use DACE (Dependency-Aware Commit Evidence) as an external evidence source. DACE provides two kinds of hints. A dependency analyzer extracts lightweight def-use and file-creation edges from the unified diff. A hunk profiler uses path and identifier similarity to suggest related hunks and warn about likely locality traps, such as same-file hunks with different roles or mixed test and implementation edits. Agents receive these outputs only as soft evidence and still choose the number of commits, hunk grouping, and replayable sequence. DACE does not report observed commit counts, reference labels, commit messages, commit hashes, intermediate states, or other information derived from git history. We mark these evidence-augmented runs with a \texttt{+T} suffix (e.g., \texttt{MiniMax+T}).

\subsection{Experimental Setup}
\label{sec:experimental_setup}

We evaluate six diagnostic baselines and four model-agent setups on the same 800 episodes. OpenAI's Codex CLI drives GPT-5.4. Claude Code CLI drives, through provider routing, ZhipuAI GLM-5, Moonshot Kimi K2.5, and MiniMax-M2.5. To keep figures and tables compact, we use the setup labels GPT-5.4, GLM-5, Kimi, and MiniMax after this point. These labels always denote model-agent setups, not standalone models. GLM-5, Kimi, and MiniMax share one harness, while the GPT-5.4 comparison also changes the CLI. Because the agent harness shapes how a model inspects files, runs commands, and formats outputs, the evaluated unit is the model-agent setup rather than the model alone. All four setups use the same task description, one prompt, and one run per episode. They can read the working tree but cannot access git history.

\subsection{Evaluation Protocol}
\label{sec:eval_protocol}

Each method is evaluated in a workspace containing the base repository snapshot at $c_0$ and the squashed diff. The workspace withholds \texttt{.git} metadata, commit hashes, commit messages, original commit boundaries, and intermediate states; traces were checked for history-inspection commands before scoring. LLM agents have at most 50 turns and may inspect the working tree or run local analysis commands. The prompt asks agents to produce a reviewable sequence of atomic commits and notes that hunks from the same file may belong to different commits. It does not provide examples from the evaluation set or specify the number of commits to produce, because predicting $m$ is part of the task. Each method outputs an ordered partition as JSON: a list of predicted commits, each with hunk indices and a short message. Messages are collected because commit cleanup normally produces them, although message quality is not scored. Some agent outputs contain coverage errors, such as missing hunk indices or duplicate assignments. For metric computation, we normalize these outputs: missing hunks are appended to a final commit, and duplicate assignments are removed after their first occurrence. Normalized scores are optimistic for malformed outputs. The strict-only check reports whether rankings survive when such outputs are discarded.

%% file: sections/evaluation.tex
\section{Results}
\label{sec:experiments}

We organize the results around three questions that separate task characterization, current capability, and error mechanisms. First, a retrospective-history benchmark is useful only if natural squashed changes expose structure that replay checks, simple grouping rules, or standard synthetic tangles do not already capture. RQ1 therefore asks what makes natural reconstruction non-trivial, using replay validity, diagnostic baselines, and matched synthetic composites. Second, RQ2 asks how current model-agent setups perform on this task. We score all 800 episodes with ARI and use TCR and TBR, difficulty slices, alternative references, and strict-output checks to interpret the scores under non-unique histories. Third, RQ3 asks what kind of decisions remain difficult and whether DACE evidence helps agents make those decisions.

\subsection{RQ1: What Makes Natural Reconstruction Non-Trivial?}
\label{sec:finding1}

\begin{figure}[H]
\centering
\includegraphics[width=0.70\linewidth]{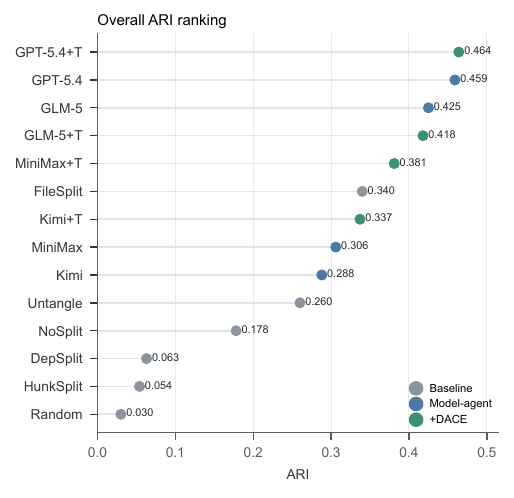}
\caption{Main grouping results on 800 natural retrospective episodes. Points report ARI against the observed developer decomposition; colors distinguish diagnostic baselines, model-agent setups, and DACE-augmented runs.}
\label{fig:main_ari}
\end{figure}

RQ1 characterizes whether natural retrospective reconstruction adds an empirical challenge beyond simpler alternatives. If replayable output were enough, if simple grouping rules recovered most developer organization, or if synthetic tangled changes exposed the same boundary structure, then existing checks would already capture much of the task. We therefore design RQ1 around three comparisons: PPAR versus ARI to separate replay from grouping, diagnostic baselines to test simple organization rules, and matched synthetic composites to test whether cross-episode tangles behave like real same-session squashed changes.

The first comparison asks whether replay validity can stand in for grouping quality. We run every baseline and model-agent setup on the same 800 natural episodes, materialize each predicted ordered history, and score each output with both PPAR and ARI. Figure~\ref{fig:main_ari} shows that replay failures are rare: all non-random methods reach PPAR $\geq 0.988$, so most predicted histories can be replayed as patch sequences. The grouping results are much less uniform. ARI spans 0.03 to 0.46, showing that different setups split the same squashed patch in substantially different ways and reach different degrees of alignment with the observed human-maintained grouping. Thus, the main challenge is not making a patch sequence apply, but recovering a maintainable grouping of change intents within that sequence.

The second comparison asks whether lightweight organization rules can stand in for reconstruction. We test five simple rules: NoSplit puts all hunks into one commit, HunkSplit puts each hunk into its own commit, FileSplit groups by changed file, DepSplit groups by dependency components, and Random provides a chance baseline. Figure~\ref{fig:main_ari} shows that these rules have clear ceilings. FileSplit is the strongest simple heuristic (ARI=0.340) because file boundaries often approximate intent boundaries, but it still misses intra-file tangling and multi-file intents. NoSplit and HunkSplit are replay-friendly extremes: one collapses every intent into a single commit, while the other separates hunks that often belong together. Random and DepSplit perform poorly, showing that ARI is not mechanically improved by producing more partitions or following dependency reachability alone. The useful structure is more specific: a decomposition must separate unrelated nearby hunks while keeping distributed edits together when they implement one change intent.

The third comparison asks whether synthetic tangled changes can substitute for real same-session squashed changes. Prior untangling work often constructs a tangled change by combining otherwise unrelated commits. If that construction had the same boundary structure as natural episodes, then synthetic composites would be a sufficient evaluation target. To test this, we build a matched synthetic set: for each real episode with $k$ observed commits, we concatenate $k$ commits drawn from distinct AtomicCommitBench episodes and use each source commit as the reference label. This keeps the commit count matched while removing the same-session constraint.

\begin{figure}[H]
\centering
\includegraphics[width=0.70\linewidth]{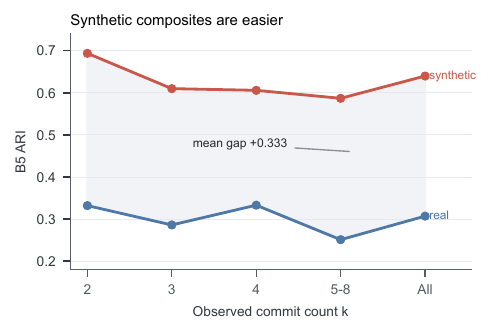}
\caption{B5 ARI on synthetic composites versus real squashed diffs, matched by observed commit count $k$. B5 receives the true $k$ for both sides; the shaded band highlights the ARI gap.}
\label{fig:synthetic_gap}
\end{figure}

We then run the same untangling baseline, B5, on the synthetic and natural sets. B5 receives the true $k$ on both sides, so the comparison tests boundary separability rather than commit-count prediction. Figure~\ref{fig:synthetic_gap} shows that synthetic composites are much easier: B5 scores an average +0.333 ARI higher on synthetic composites than on real squashed diffs ($p<10^{-49}$, Cliff's $\delta=0.430$). The reason is that cross-episode construction often combines commits from different files, identifiers, or topics, making their boundaries clearer. Real same-session changes are harder because the commits come from one development context: they may touch the same files, share helper edits, or modify related logic. Natural episodes therefore cannot be replaced by simple synthetic composites; they measure boundary ambiguity that arises inside real development sessions.

\findingbox{Observation~RQ1}{Natural retrospective episodes expose grouping difficulty that replay checks, simple heuristics, and synthetic composites miss. Non-random methods nearly all replay (PPAR $\geq 0.988$), yet grouping quality still varies widely (0.03--0.46 ARI), and matched synthetic composites are much easier than real squashed diffs (+0.333 ARI).}

\subsection{RQ2: How Do Current Model-Agent Setups Perform?}
\label{sec:finding2}

RQ2 asks whether current model-agent setups can turn one large patch into a decomposition that resembles how a human maintainer organized the same change. Since several histories may be defensible, the target is human-aligned draft history rather than exact boundary recovery. We assess this capability with complementary signals. ARI measures alignment with an observed human-maintained grouping. TCR and TBR check whether predicted commits localize behavior changes on the scoreable test slice. Difficulty, alternative-reference, and strict-output checks test how stable the pattern is across harder episodes, different reference decompositions, and parser-normalization choices.

The main ARI results show that model-agent setups recover substantial human-aligned structure from squashed patches. In Figure~\ref{fig:main_ari}, all four model-agent setups score far above Random and the two replayable extremes, indicating meaningful grouping signal. The GPT-5.4 setup reaches ARI=0.459 and the GLM-5 setup reaches 0.425, both above FileSplit, the strongest simple baseline (0.340). FileSplit is a meaningful diagnostic target because file paths often approximate developer intent. The MiniMax setup (0.306) and Kimi setup (0.288) are close to this strong path-local heuristic, while the GPT-5.4 and GLM-5 setups move further toward the organization seen in human-maintained histories. Overall, current model-agent setups can produce plausible commit decompositions for a large patch, and stronger setups produce decompositions that are measurably closer to human-maintained organization than file-local grouping alone. The practical reading is that these outputs are structured draft histories: they provide replayable commit units that expose candidate review units for a maintainer or later agent to inspect and refine. The final code is fixed across methods; the scores compare history organization. GLM-5, MiniMax, and Kimi share one harness; the GPT-5.4 comparison also includes the Codex CLI harness, so we interpret the rows as model-agent setups rather than isolated model rankings.

Strict output coverage is also an operational signal. Before normalization, strict valid counts are 798 of 800 for GPT-5.4, 712 of 800 for GLM-5, 652 of 800 for MiniMax, and 550 of 800 for Kimi. These rates show that producing a complete, non-duplicated hunk assignment is itself uneven across setups. Discarding coverage-error outputs before scoring preserves the same broad ranking, so the ARI pattern is not an artifact of optimistic normalization.

Commit-count prediction explains only part of this level. Under-splitting and over-splitting both hurt ARI, but many errors occur at fixed or near-correct commit counts. The harder decision is assigning hunks to change intents; estimating $m$ is one part of the reconstruction.

TCR adds a behavior-containment signal to the reference-based ARI result. It is computed where modified tests run at the episode tip and partial reverse patches yield a scoreable attribution surface; episodes outside this slice remain in the 800-episode PPAR and ARI evaluation. On the 151-episode common scoreable slice, TCR ranks the setups in the same order as ARI: GPT-5.4 (0.917), GLM-5 (0.871), MiniMax (0.827), and Kimi (0.780), with the observed developer history at 0.922 (Figure~\ref{fig:tcr}). The slice is harder than the rest of the benchmark, and every setup's ARI is lower on it, making the agreement a useful partial stress check rather than a full correctness result.

\begin{center}
\begin{minipage}{0.48\linewidth}
\centering
\includegraphics[width=\linewidth]{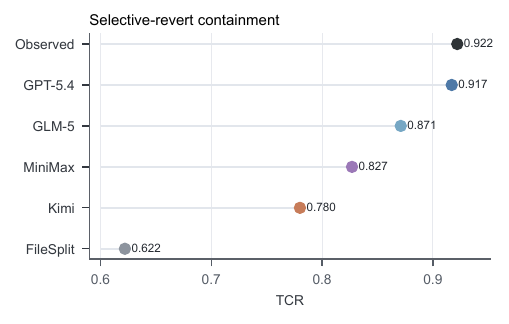}
\captionof{figure}{Selective-revert containment on the 151-episode common scoreable slice.}
\label{fig:tcr}
\end{minipage}\hfill
\begin{minipage}{0.48\linewidth}
\centering
\includegraphics[width=\linewidth]{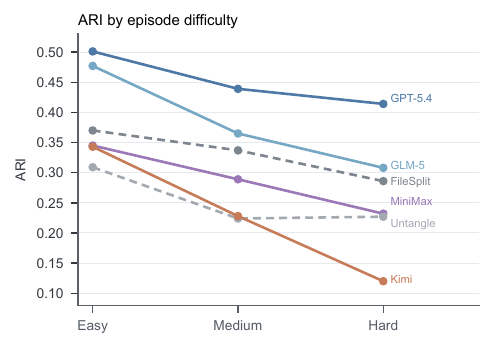}
\captionof{figure}{ARI by difficulty tier. Difficulty is based on commit count and diff size.}
\label{fig:difficulty}
\end{minipage}
\end{center}

This probe clarifies why grouping matters for maintenance. If unrelated changes are lumped together, one reverse patch can remove both behaviors and mix their test failures. If one logical change is split across commits, either fragment can break tests whose cause is distributed across the sequence. High TCR means the predicted commits behave more like isolated failure-containment units under this patch-based test. B0 shows why containment must be read with TBR: its one-commit containment is vacuous, and it has the highest TBR (0.094), compared with 0.043 for GPT-5.4 and 0.042 for FileSplit.

Capability is strongest on easier structure and degrades on harder episodes. The gap widens as episodes become harder: the GPT-5.4 setup holds ARI=0.414 on Hard episodes, while the Kimi setup drops to 0.120 (Figure~\ref{fig:difficulty}). To check that this pattern is not an artifact of binning by commit count and diff size, we re-partition difficulty by dependency-graph width, longest dependency chain, and intra-file split count. Under each feature, the gap between the GPT-5.4 and Kimi setups widens across tertiles, from under 0.07 on the easiest third to about 0.2 on the hardest. FileSplit remains relatively stable because its rule does not change with diff size; it fails when change intents cross files or split within a file. Setup degradation is more gradual, suggesting that the evaluated model-agent systems use more than file-level or count-level heuristics.

\begin{table}[t]
\centering
\caption{Reference-sensitivity check. Rows and columns use the setup labels defined in Section~\ref{sec:experimental_setup}; cells report mean ARI against each alternative reference.}
\label{tab:ref_stability}
\small
\begin{tabular}{@{}lccccc@{}}
\toprule
Setup & Observed & GPT-5.4 & GLM-5 & Kimi & MiniMax \\
\midrule
GPT-5.4 & 0.459 & --- & 0.591 & 0.446 & 0.454 \\
GLM-5 & 0.425 & 0.591 & --- & 0.496 & 0.498 \\
Kimi & 0.288 & 0.446 & 0.496 & --- & 0.444 \\
MiniMax & 0.306 & 0.454 & 0.498 & 0.444 & --- \\
\bottomrule
\end{tabular}
\end{table}

Alternative references preserve the same pattern. We rescore each setup against alternative setup decompositions, and the GPT-5.4 and GLM-5 setups remain above MiniMax and Kimi under every external reference (Table~\ref{tab:ref_stability}). This check captures robustness across plausible stylistic variation. Episode-level bootstrap 95\% confidence intervals over mean ARI on this fixed benchmark (10{,}000 resamples) separate the same two performance bands: GPT-5.4 [0.41, 0.49] and GLM-5 [0.40, 0.47] do not overlap with Kimi [0.28, 0.35] or MiniMax [0.31, 0.37]. Pairwise Wilcoxon signed-rank tests~\cite{wilcoxon1945} with Benjamini-Hochberg correction~\cite{benjamini1995fdr} give the same cross-band separation, with every cross-band pairing at $p_\text{BH}<10^{-10}$. The most stable conclusion is the cross-band separation; within-band ordering is weaker.

Additional checks preserve the same interpretation. Restricting to the 356 episodes whose hunks are all cleanly attributable to a single observed commit leaves the setup ranking unchanged. Since all 10 repositories are public, some historical commits may have appeared in model pretraining corpora. We therefore split results by each model's public pretraining cutoff. Post-cutoff means are higher for all four setups: GPT-5.4 +0.163, MiniMax +0.080, GLM-5 +0.059, and Kimi +0.007. This split points away from training-period memorization as the main driver.

\findingbox{Observation~RQ2}{Current model-agent setups can reconstruct useful draft histories from squashed patches. All four setups recover more structure than degenerate baselines; GPT-5.4 (0.46 ARI) and GLM-5 (0.43) outperform the strongest simple heuristic, while TCR, alternative-reference, and strict-output checks preserve the same broad performance bands.}

\subsection{RQ3: What Explains the Remaining Gap?}
\label{sec:finding3}

RQ3 moves from aggregate scores to the decisions that remain difficult. The RQ2 results suggest that locality is an important candidate mechanism: FileSplit is competitive, yet it has a clear ceiling, and harder episodes increase the gap between setups. We therefore use qualitative trace inspection to identify recurring error patterns, then compare base runs with DACE-augmented runs to test whether explicit evidence changes aggregate scores. DACE (Dependency-Aware Commit Evidence) supplies soft evidence from a dependency analyzer and a hunk profiler; agents still choose the number of commits, hunk grouping, and replayable sequence.

The trace analysis points to a locality tension. Nearby hunks often belong together, but maintainable histories also require splitting nearby hunks and attaching distributed support edits to the right change intent. In the inspected traces, the clearest recurring patterns are same-file lumping, where file locality overrides the semantic role of each hunk, and support-hunk drift, where tests or helper edits follow the nearest textual change.

\begin{figure}[H]
\centering
\includegraphics[width=0.95\linewidth]{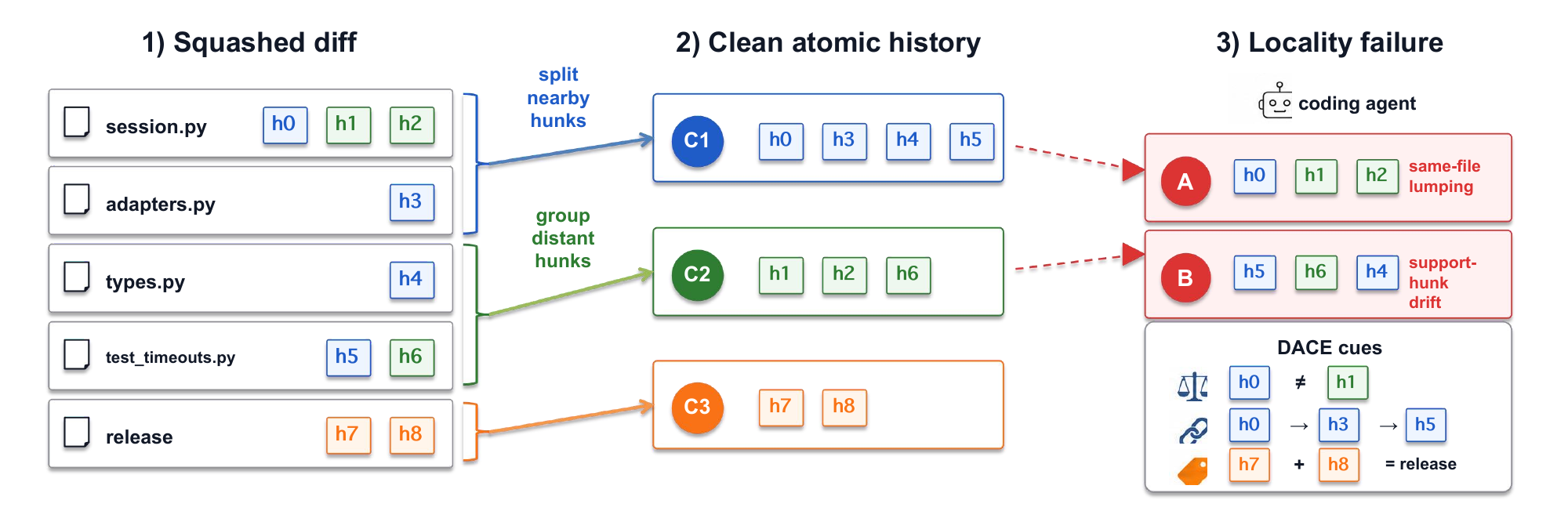}
\caption{An anonymized representative reconstruction case. The squashed diff interleaves a timeout bug fix, an option-parser refactor, and release metadata. A maintainable history must split same-file hunks, group cross-file hunks, and keep support hunks with the change intent they support.}
\label{fig:case_study}
\end{figure}

Figure~\ref{fig:case_study} illustrates why this tension matters. The squashed diff contains a timeout bug fix, an option-parser refactor, and release metadata. The observed reference history groups the timeout-related hunks across source, type, and test files. It separates nearby same-file refactoring hunks and keeps the changelog with the version bump. File-level grouping fails in both directions: it lumps the timeout-fix hunk with same-file refactoring hunks in \texttt{session.py}, while separating cross-file hunks that jointly implement the timeout fix. A locality-driven agent can make the analogous error by attaching support hunks to the nearest textual edit rather than to the change intent they support.

The DACE comparison then tests whether explicit evidence helps with this locality tension. If a setup over-relies on nearby paths or same-file grouping, dependency and hunk-role cues should help it separate nearby but unrelated edits and connect distributed support hunks. The results follow this pattern: adding the dependency analyzer and hunk profiler raises MiniMax from 0.306 to 0.381 ARI (+0.075) and Kimi from 0.288 to 0.337 (+0.049), with paired Wilcoxon tests~\cite{wilcoxon1945} and Benjamini-Hochberg correction~\cite{benjamini1995fdr} giving $p<10^{-8}$ and $p<10^{-6}$, respectively. For the two higher-scoring setups, the same evidence source leaves scores statistically unchanged: GLM-5 moves by $-0.006$ ($p=0.56$) and GPT-5.4 by $+0.005$ ($p=0.90$). The benefit is therefore targeted rather than universal.

The inspected traces give a concrete reading of the aggregate gains. DACE does not choose commits for the agent; it exposes dependency cues from lightweight edges and grouping cues from hunk similarity and change-role warnings. A typical successful use is anti-lumping: when same-file hunks have different change roles, the hunk profiler makes that local evidence explicit, helping the agent separate a behavior change from a nearby refactor or cleanup. The aggregate gains fit this interpretation. Tool augmentation helps most when path-based grouping is plausible but wrong, such as same-file hunks with different roles or cross-file helper edits tied together through imports. It helps less when the decision requires broader design judgment.

\findingbox{Observation~RQ3}{The remaining errors concentrate around locality: agents often lump unrelated same-file hunks or attach support hunks to the nearest edit. DACE evidence helps the lower-scoring setups, MiniMax (+0.075 ARI) and Kimi (+0.049), mainly by making anti-lumping and dependency cues explicit.}

%% file: sections/related.tex
\section{Related Work}
\label{sec:related}

\subsection{Coding-Agent Benchmarks and Process-Oriented Evaluation}

SWE-bench~\cite{jimenez2024swebench} and its variants, including Verified~\cite{chowdhury2024swebenchverified}, Multi-SWE-bench~\cite{zan2025multiswebench}, and SWE-bench-Live~\cite{zhang2025swebenchlive}, evaluate whether a coding agent can produce a patch that passes a repository's tests. Systems such as SWE-Agent~\cite{yang2024sweagent} and AutoCodeRover~\cite{zhang2024autocoderover} have driven progress on final-state repair. AtomicCommitBench asks a different artifact question: after a multi-commit change already exists, can an agent reconstruct a reviewable, replayable commit sequence?

Recent work has begun to evaluate process-level aspects of coding agents. GitGoodBench~\cite{joshi2025gitgoodbench} evaluates agentic Git operations across file-commit, issue-resolution, and issue-classification scenarios, using LLM-as-Judge for the file-commit-chain setting; EditFlow~\cite{sun2026editflow} reconstructs developer editing flows; SWE-EVO~\cite{fan2025sweevo} evaluates long-horizon evolution; and AgentPack~\cite{zi2025agentpack} collects agent-human co-authored edits with structured commit messages. Pham and Ghaleb~\cite{pham2026agentpr} compare structural properties of AI-generated and human pull requests. None formalizes retrospective commit-history reconstruction as hunk-level commit decomposition with replay validity and deterministic grouping metrics.

Another line studies how agents consume commit history. Code Researcher~\cite{singh2025coderesearcher} uses commit history for repository search, HAFixAgent~\cite{shi2025hafixagent} conditions repair on commit-level context, and Lore~\cite{stetsenko2026lore} treats commit messages as a structured knowledge protocol. These systems motivate the downstream value of organized histories: commits record work for humans and can serve as retrieval units for future agents. Our work measures the producer side of that loop.

\subsection{Commit Quality and Tangled Changes}

Prior SE research shows that commit structure has measurable consequences. Tangled changes can inflate defect prediction error by 5 to 200\%~\cite{herzig2016impact}; tangling has been studied in fine-grained bug-fix datasets~\cite{herbold2022tangling} and as a multi-label concern-detection task~\cite{koh2026detecting}. Program-dependence-based changeset decomposition supports review~\cite{barnett2015decomposition}, and controlled evidence suggests decomposed changesets reduce wrongly reported issues and context seeking during review~\cite{dibiase2018change}. Code-review research similarly treats reviewability as an artifact property~\cite{rigby2014peer,macleod2018code}.
These studies show that atomic histories matter, but they do not evaluate coding agents as producers of commit histories. AtomicCommitBench turns retrospective cleanup into an evaluation target: agents must organize hunks into a replayable history after the final tree already exists.

Commit message generation is complementary. Datasets such as CommitBench~\cite{schall2024commitbench} and CommitChronicle~\cite{eliseeva2023commitchronicle} evaluate whether systems can produce messages from existing diffs, and RAG or reasoning-based methods improve message quality~\cite{li2024react,wang2024omg}. Our task asks how the diff should be grouped into commits before messages are written. We collect short messages because cleanup normally produces them, although message quality is not scored.

\subsection{Commit Untangling}

Commit untangling is the closest prior task. SmartCommit~\cite{shen2021smartcommit} constructs a change graph and partitions it into activity-oriented groups. Flexeme~\cite{barnett2020flexeme} overlays program-flow information on diffs and applies spectral clustering. UTango~\cite{wang2022utango} learns context-aware graph embeddings for untangling. LLM-based systems such as Atomizer~\cite{wu2026atomizer} and ColaUntangle~\cite{hou2025colautangle} use intent reasoning or collaborative agents to identify change groups.

These systems provide the closest partitioning tradition. AtomicCommitBench builds on it but evaluates a commit-history artifact: hunks must form coherent commits, and the resulting sequence must replay. We use lightweight heuristic baselines as diagnostic references; adapting full untangling systems with commit-count selection and replay repair is a separate comparison.

Final-patch benchmarks score the final tree but produce no partition. Process-oriented benchmarks evaluate ordered behavior yet still emit no hunk partition. Untangling systems partition multi-file changes but stop at change-intent groups rather than replayable commit histories. Retrospective commit-history reconstruction combines hunk grouping, commit-count selection, replay validity, and evaluation on real multi-file changes.

%% file: sections/discussion.tex
\section{Discussion}
\label{sec:discussion}

Coding-agent evaluation today asks whether the final patch is correct. AtomicCommitBench asks whether the delivered change is organized into a history the next reader can use. A monolithic commit and a clean commit sequence can carry identical code while placing different burdens on review, selective revert, and history-based retrieval. Our results show that this property needs separate measurement: replay validity nearly saturates, synthetic composites are easier than natural same-session diffs, and current agents can produce draft histories with meaningful alignment to human-maintained organization.

\subsection{Agents as Producers and Consumers of History}

Agents now sit on both ends of the commit-history loop. They produce history through session-level squashed changes, and they consume it when localizing bugs or retrieving prior edits~\cite{singh2025coderesearcher,shi2025hafixagent,stetsenko2026lore}. The history one agent leaves behind can become context another agent reads, so poor organization may become degraded input to later automation.

AtomicCommitBench measures the producer side: can current agents organize a completed change well? We leave the consumer-side question open. Better-organized history may help downstream agents retrieve prior edits or repair bugs, but this paper does not test that causal link.

\subsection{A Reusable Measurement Structure}

The measurement structure also applies beyond commit history. Many agentic-coding artifacts have no single correct answer, yet still have a cheap deterministic validity check. For such tasks, the same pattern can work: require the validity gate, use a reference-based relative score, and add a behavioral check when available.

AtomicCommitBench is most useful as a comparative diagnostic. It can test whether a new prompt, model, or tool reduces same-file lumping and support-hunk drift while preserving replay validity and selective-revert containment. The design implication is narrower than a leaderboard: support tools should expose structured evidence, such as hunk roles and dependency edges, while leaving final grouping to the agent. Commit-history reconstruction gives agentic-coding evaluation a concrete way to assess the artifacts an agent leaves behind alongside the code it produces.

%% file: sections/threats.tex
\section{Threats to Validity}
\label{sec:threats}

\noindent\textbf{Task and metric validity.}
AtomicCommitBench evaluates retrospective decomposition: reorganizing an existing squashed change into a commit sequence. Prospective commit authoring asks what an agent should stage while editing, testing, and revising code. The retrospective setting isolates history organization from patch construction and matches cleanup workflows such as \texttt{git rebase -i}, squash-then-split review cleanup, and hunk-level staging.
Besides, the hunk-level representation is imperfect: a content-based audit shows that 2.3\% of squashed hunks contain content attributable to more than one observed commit. We keep this representation because hunk-level staging is the native cleanup granularity and setup rankings are unchanged on the fully clean subset. Sub-hunk decomposition remains outside this paper. Semantic coherence also has no unique ground truth, so ARI is a dataset-level relative score rather than an absolute per-episode correctness score; we check reference sensitivity with alternative setup decompositions.
PPAR and TCR cover different parts of the construct. PPAR is a structural gate: a replayable sequence can still group intents poorly, and a non-replayable sequence can fail for formatting or patch-application reasons. TCR provides a label-free containment check only where modified-test execution yields an attribution signal. It applies selected hunks from a reverse diff with \texttt{git apply}; it does not evaluate full \texttt{git revert} behavior, commit metadata, merge semantics, unmodified regression tests, or non-test behavior. The common slice for setup outputs and the observed history contains 151 such episodes, so TCR is an independent partial probe rather than a full-benchmark correctness score.

\noindent\textbf{Dataset and external validity.}
The evaluation set is drawn from 10 mature Python repositories. These projects provide maintainer-accepted public histories rather than adjudicated atomicity labels, and they do not represent all software. Results may differ for enterprise repositories, less curated projects, other languages, or ecosystems with different build systems. The same-author and 48-hour window constraints bias the dataset toward coherent cleanup episodes, matching the intended retrospective setting but not the full range of collaborative development.
All repositories are public, so pretraining exposure is possible. Splitting results by each model's public pretraining cutoff shows higher post-cutoff mean ARI for all four setups, which is not the pattern expected under simple memorization. We cannot verify model-specific cutoff claims or fully rule out exposure; the check is bounded evidence, not proof of absence.

Commit untangling systems are the closest algorithmic baselines~\cite{shen2021smartcommit,barnett2020flexeme,wang2022utango,wu2026atomizer}. We use B5 as a transparent diagnostic anchor; applying full untangling systems to AtomicCommitBench would require commit-count selection and replay-validation components. This keeps the distinction between partitioning alone and replayable commit-history reconstruction.

\noindent\textbf{Model-agent setup and evidence-source validity.}
The evaluated setups use heterogeneous harnesses: Codex CLI with GPT-5.4 for the GPT-5.4-labeled setup and Claude Code CLI with provider routing for GLM-5, Kimi, and MiniMax. The same-harness comparison among GLM-5, Kimi, and MiniMax is compute-matched within 7\% by mean token usage; the comparison between GPT-5.4 and GLM-5 is a setup-level comparison. Each episode is evaluated with one run under one prompt, producing a reproducible snapshot of these hosted setups. The confidence intervals and $p$-values are episode-level descriptive analyses conditional on fixed traces, not cluster-robust population inference.

The +T setting uses DACE as an evidence source, not as an oracle. The tools report dependency edges, similarity clusters, and change-role warnings, but not observed commit counts, reference labels, commit messages, commit hashes, or intermediate states. We evaluate the combined tool package without dependency-only or profiler-only ablations. The observed gains support this evidence source under the chosen prompt, not a claim about which component is causal.
The strongest claims concern the task formalization, benchmark construction, and separation of replay, grouping, and selective-revert measurements. The setup ranking and tool effects diagnose the evaluated setups rather than establishing a permanent leaderboard ordering.

%% file: sections/conclusion.tex
\section{Conclusion}
\label{sec:conclusion}

We studied retrospective commit-history reconstruction: organizing a completed squashed change into a replayable sequence of commits. We formalized the task as hunk-to-commit partitioning with a replay requirement and built AtomicCommitBench from 800 real multi-commit episodes across 10 Python projects. The benchmark combines PPAR for replay, ARI for reference-based grouping, and TCR for failure containment on a selected modified-test slice. Natural episodes are necessary: replay validity nearly saturates, grouping quality varies widely, and matched synthetic composites are much easier than real same-author squashed diffs. In this model-agent snapshot, the GPT-5.4 and GLM-5 setups recover more structure than a strong file-based heuristic, while the MiniMax and Kimi setups remain closer to path-local organization. Qualitative trace diagnosis identifies same-file lumping and support-hunk drift as recurring error patterns, and DACE helps mainly lower-scoring setups under this evidence source and prompt by exposing dependency and hunk-role cues. Agentic-coding evaluation should measure the history an agent leaves behind alongside the code it produces.